\documentclass[useAMS,usenatbib]{mn2e}

\usepackage[pdftex]{graphicx}

\def \aap {A\&A }
\def \apj {ApJ }
\def \apjs {ApJS }
\def \apjl {ApJL }

\def \mnras {MNRAS }
\def \aj {AJ }

\def \prd {PhRvD }
\def \apss{Ap\&SS }
\def \aaps{A\&AS }

\def \etal {\textit{et al. }}

\title[KIC 1571511B: A Benchmark Low-Mass Star]{KIC 1571511B: A Benchmark Low-Mass Star In An Eclipsing Binary System In The {\it Kepler} Field\thanks{Based on observations made with the Nordic Optical Telescope, operated on the island of La Palma jointly by Denmark, Finland, Iceland,Norway, and Sweden, in the Spanish Observatorio del Roque de los Muchachos of the Instituto de Astrofisica de Canarias}}
\author[A. Ofir]{A.Ofir$^{1}$\thanks{E-mail: avivofir@wise.tau.ac.il}, D. Gandolfi $^{2,4}$, Lars Buchhave$^{5,6}$,  C. H. S. Lacy$^{3}$, A. P. Hatzes$^{4}$ and Malcolm Fridlund$^{2}$ \\
$^{1}$School of Physics and Astronomy, Raymond and Beverly Sackler Faculty of Exact Sciences, Tel Aviv University, Tel Aviv, Israel\\
$^{2}$Research and Scientific Support Department, ESTEC/ESA, PO Box 299, 2200 AG Noordwijk, The Netherlands\\
$^{3}$Department of Physics, University of Arkansas, Fayetteville, AR 72701, USA\\
$^{4}$Th\"uringer Landessternwarte, Sternwarte 5, Tautenburg, D-07778 Tautenburg, Germany\\
$^{5}$Harvard-Smithsonian Center for Astrophysics, Cambridge, USA\\
$^{6}$Niels Bohr Institute, Copenhagen University, Denmark\\
}

\begin{document}

\date{Submitted...}

\pagerange{\pageref{firstpage}--\pageref{lastpage}} \pubyear{2010}

\maketitle

\label{firstpage}

\begin{abstract}
KIC 1571511 is a 14d eclipsing binary (EB) in the {\it Kepler} dataset. The secondary of this EB is a very low mass star with a mass of $0.14136 \pm 0.00036 M_\odot$ and a radius of $0.17831 _{-0.00062}^{+0.00051} R_\odot$ (statistical errors only). The overall system parameters make KIC 1571511B an ideal "benchmark object": among the smallest, lightest and best-described stars known, smaller even than some known exoplanet. Currently available photometry encompasses only a small part of the total: future {\it Kepler} data releases promise to constrain many of the properties of KIC 1571511B to unprecedented level. However, as in many spectroscopic single-lined systems, the current error budget is dominated by the modeling errors of the primary and not by the above statistical errors. We conclude that detecting the RV signal of the secondary component is crucial to achieving the full potential of this possible benchmark object for the study of low mass stars. 
\end{abstract}


\section{Introduction}
\label{Intro}
Current stellar models describe well the basic properties of a wide range of stars. However, there are persisting discrepancies at the lower end of the stellar mass range, where very late-type stars have measured radii that are higher than theory predicts (e.g. Lacy 1977, Ribas 2006, Torres \& Ribas 2002). This letter reports the discovery that {\it Kepler} target KIC 1571511 is an EB which contains a very low-mass star as secondary -- hereafter just 511B -- based on the public Q0-Q2 {\it Kepler} data and our radial velocity (RV) follow-up. We point out the special location of 511B in the parameter space and offer to use it as a benchmark object for future studies of low-mass stellar objects. In \S \ref{IdentAndObs} we highlight the important properties of the KIC 1571511 photometry and subsequent RV follow-up, in \S \ref{Modeling} we model the system, in \S \ref{PysPar} we estimate the system's physical parameters based on all available data, and conclude in \S \ref{Discussion}.

\section{Identification and Follow-up}
\label{IdentAndObs}

We identified KIC 1571511 as an interesting system as early as four days after the {\it Kepler} Q0-Q1 data was made public [Borucki \& the Kepler Team 2010, see references therein for a description of the {\it Kepler} satellite and data]. This system contains periodic eclipse-like events every 14.02d, with a depth of $\sim 2\%$ - a conspicuous signal at {\it Kepler}'s high precision. At that time we suspected this system to host a giant transiting planet overlooked by the {\it Kepler} team due to its (assumed) orbital eccentricity. It is noteworthy that while high-eccentricity transiting planets can generate transits that are "too" long and photometrically identical to EBs on a circular orbit, the {\it Kepler} pipeline does not consider eccentric orbits in its Data Validation module \footnote{Kepler Data Processing Handbook \S 9.3, document number KSCI-19081-001 of 1 April 2011.}. We thus were able to secure a few RV measurements with the FIES spectrograph at the NOT observatory to test the giant-eccentric-planet hypothesis. We note that this object was also considered as an overlooked planetary candidate by Coughlin \etal (2010) but they did not provide RV data. 

RV follow-up of KIC 1571511 was started in October 2010 using the FIES fiber-fed \'Echelle spectrograph attached to the 2.61\,m Nordic Optical Telescope (observing program P41-426). The observations were performed under grey/dark time with good and stable sky-conditions. The 1.3 $\mu m$ high-resolution fiber was employed, yielding a resolving power of $\lambda/\Delta \lambda\approx 67,000$ and a wavelength coverage of about 3600 - 7400 \AA.  Following the method described in Buchhave \etal (2010), long-exposed ThAr spectra were acquired immediately before and after each target spectrum to improve the wavelength solution and trace any instrumental drifts. Standard IRAF routines were used for the data reduction and spectra extraction. The RV measurements were derived by cross-correlating the target spectra with a spectrum of the RV standard star HD182488 (Udry \etal , 1999) observed with the same instrument set-up as the target. A journal of the FIES observations is given in Table {\ref{RVvalues}. The FIES observations revealed a single-line spectroscopic binary (SB1) with an eccentric orbit ($e \cong 0.33$) and a RV semi-amplitude of $K \cong 10.5$ km/s, compatible with a very low-mass companion star orbiting the main component. Figure \ref{ModelFigure} (right-hand panels) show the FIES RV measurements along with the Keplerian RV curve resulting from the best-fitting simultaneous photometric-RV solution (see \S \ref{Modeling}) and residuals.

\begin{table}
\caption{Measured Radial Velocities}
\begin{tabular}{l l l l l} 
\hline
\label{RVvalues}
HJD 				& RV               & RV error         & $T_{\rm{exp}}$ & S/N per pixel   \\
-2,450,000   	& [km s$^{-1}$] & [km s$^{-1}$] & [s]   &   at 6000 \AA \\
\hline
5479.36633498  & -29.998  & 0.046 & 2700   &   26 \\
5482.46744505  & -10.181  & 0.042 & 2400   &   22 \\
5484.50488456  & -13.196  & 0.073 & 1500   &   12 \\
5491.37000002  & -29.871  & 0.054 & 1200   &   14 \\
5497.32650926  & -10.578  & 0.055 & 1200   &   16 \\
5518.32931804  & -27.713  & 0.060 & 1200   &   16 \\
\hline
\end{tabular}
\end{table}

\section{System modeling with MCMC}
\label{Modeling}

\subsection{Pre-processing}
Since both the primary and secondary eclipses were easily identified using a weak filter we were able to use a stronger filtering scheme for the final light curve (hereafter LC) without modifying the eclipses themselves. We computed an iterative $\sim 1$ day long, second-order Savitzky-Golay filter with $5 \sigma$ outlier rejection using only the out-of-eclipses sections of the data, and separately for each continuous section of the Q0-Q2 data. We then interpolated the filter to the times of the eclipses and normalized the LC with that filter. No data points were rejected to this point, but only removed from the filter calculation, and this LC contained 6177 data points. Continuous sections were taken between quarters and anomalies, as reported by the {\it Kepler} Release Notes (We considered the following anomalies: attitude tweak, safe mode, Earth point, and coarse point). We note that such a strong filter automatically removes any in-phase out of eclipse variation, such as the reflection effect or the Doppler boosting signal (Zucker \etal 2007). We chose a strong filter to better remove the stellar activity since it has higher amplitude than the the Doppler boosting signal and since anyhow no further information is expected from it given our high precision RV data. Once an initial solution was obtained (below) we rejected 22 points that deviated by more than $4 \sigma$ from the model, and this data was used for the light-curve part of the final solution.

\subsection{Methods}
We used MCMC (Monte Carlo Markov Chains, e.g. Tegmark \etal 2004] for the simultaneous solution of the photometry and RV data. As in many MCMC codes, an MCMC chain begins by computing the model for a given set of parameters. We then add a random perturbation to each parameter $p$ chosen from normal distribution of width $\sigma_p$. If the total $\chi^2$ has reduced at the new perturbed location, the perturbed parameters set is accepted as the new set. If the total $\chi^2$ has increased, the new set is only sometimes accepted -- at a probability of $exp(\frac{\chi^2_{\mathrm{old}}-\chi^2_{\mathrm{new}}}{2})$ , also commonly referred to as the Hastings-Metropolis jump condition. Early chains used a rather large $\sigma_p$, or step size, as the parameters space was explored for interesting regions of low $\chi^2$. Once an initial solution was obtained we set all the different jump sizes to the standard deviation of that MCMC test-chain, and re-scaled them down by a common factor of $N_{DOF}^{1/2}$ where $N_{DOF}=13$ is the number of degrees of freedom. This is done since once all the parameters' jumps are measured in units of their own standard deviation, the MCMC process is actually a random walk in a $N_{DOF}$-dimensional space, and so the typical distance covered by such a step is $N_{DOF}^{1/2}$.

We use the above MCMC procedure and the Mandel and Agol (2002) formalism to model the system as two luminous spheres (vs. Roche geometry models). We expand the formalism to account for the secondary's flux and {\it Kepler}'s finite integration time (Kipping 2010). Specifically, eq. 40 of Kipping (2010) implied a sub-sampling of $N \simeq 2.3$ sub-samples to reduce the modeling errors to below the measurement errors. We therefore chose $N=5$ to make sure this effect is indeed minimized.

Our model included eccentric Keplerian orbits with a period $P$, a unit semi major axis $a \equiv 1$, and eccentricity and argument of periastron given by $e cos (\omega)$ and $e sin (\omega)$. The fitted reference time parameter is the more easily (and more accurately) observed time of mid eclipse $T_{mid}$ while the time of periastron passage is computed from it using the previous parameters. Other parameters are the fractional radius of the primary $r_1/a$, the relative radius of the secondary $r_2/r_1$ and the orbital inclination to the line of sight $i$. Using the Mandel and Agol (2002) limb-darkening model we inluded a quadratic limb-darkening for the primary using $u_{1,1}, u_{2,1}$ and a linear limb-darkening for the secondary with $u_{1,2}$. Since the total flux is normalized to unity, we only varied the secondary fractional luminosity $L_2/L_{tot}$. When contamination, or ``third light" $L_3$, was included contamination levels of more than $\sim 3\%$ gave poorer fits, and contamination levels $<3\% $ were just as good as the fit with zero contamination, so we adopted a fixed $L_3=0$. Finally, the mass ratio $q=\frac{m2}{m1+m2}$ affects the positions of the two bodies - and so it is included in the LC model too, and not just in the RV model. From the above it follows that for the RV model we only needed to add the systemic velocity $\gamma$ and the overall scale of the system -- how many metres there are in one semi major axis $a$. These are only a shift- and scaling- terms for the otherwise known RV morphology (from the previous parameters). In principle, the scaling parameter should be an MCMC variable and the systemic velocity should be fitted analytically - but since this is a single-lined binary the scale is degenerate with the mass ratio. We therefore did not vary the scaling parameter but fixed at a value computed from the estimated mass of the primary (\S \ref{PysPar}), the above period and mass ratio, and Kepler's laws.

\subsection{Application to KIC 1571511}

We searched the above parameter space using numerous MCMC chains until we got very close (in retrospect -- within $\Delta \chi^2 <5$) of the absolute $\chi^2$ minimum. We then ran 30 chains of $5 \times 10^4$ steps each to densely sample the local volume. We verified that all of them converged on the same parameters set and then concatenated all the chains (as they are independent, see Tegmark \etal 2004) to a final $1.5 \times 10^6$ steps long chain used for the parameters value and errors estimation. We also checked that all parameters are well mixed, i.e., their effective length (which is their nominal length divided by their autocorrelation length) is $\gg 1$. Since we started all the runs from a point very near the final minimum no ``burn-in'' was required and all steps were kept. To further check that no farther and deeper minima exist, we also ran 300 shorter chains that did not start from near the reported $\chi^2_{min}$ - but randomly perturbed by up to $100 \sigma$ in every parameter. No other deeper minimum was found. The best-fitting parameters and their errors were determined as the median values of each parameter's distribution and the the ranges that span $\frac{68.3\%}{2}$ of the steps on either side of that median. These parameters and errors, as well as some derived quantities, are reported in Table \ref{ModelResults} and the LC and RV models are shown on Figure \ref{ModelFigure}.

Overall the quality of the fit is satisfactory, with $\chi^2_{tot}$ of 8046.5, or reduced $\chi^2_{red}$ of 1.31. Some of the excess residuals can be attributed to imperfect filtering of microactivity on primary: the EB is active with an amplitude of $7.2 \times 10^{-4}$ while the LC's residuals are $1.8 \times 10^{-4}$ -- almost 4 times lower. It is unlikely that this variability is dominated by 511B since this would imply a variability of about $25\%$ in its flux on $\sim$ day time scales, and so we attribute it almost entirely to the primary (hereafter just 511A). It is noteworthy that while both limb-darkening coefficients of 511A were constrained by the data, no such constraint is yet possible even for the simplest linear model of 511B.

\begin{figure*}
\includegraphics[width=177mm]{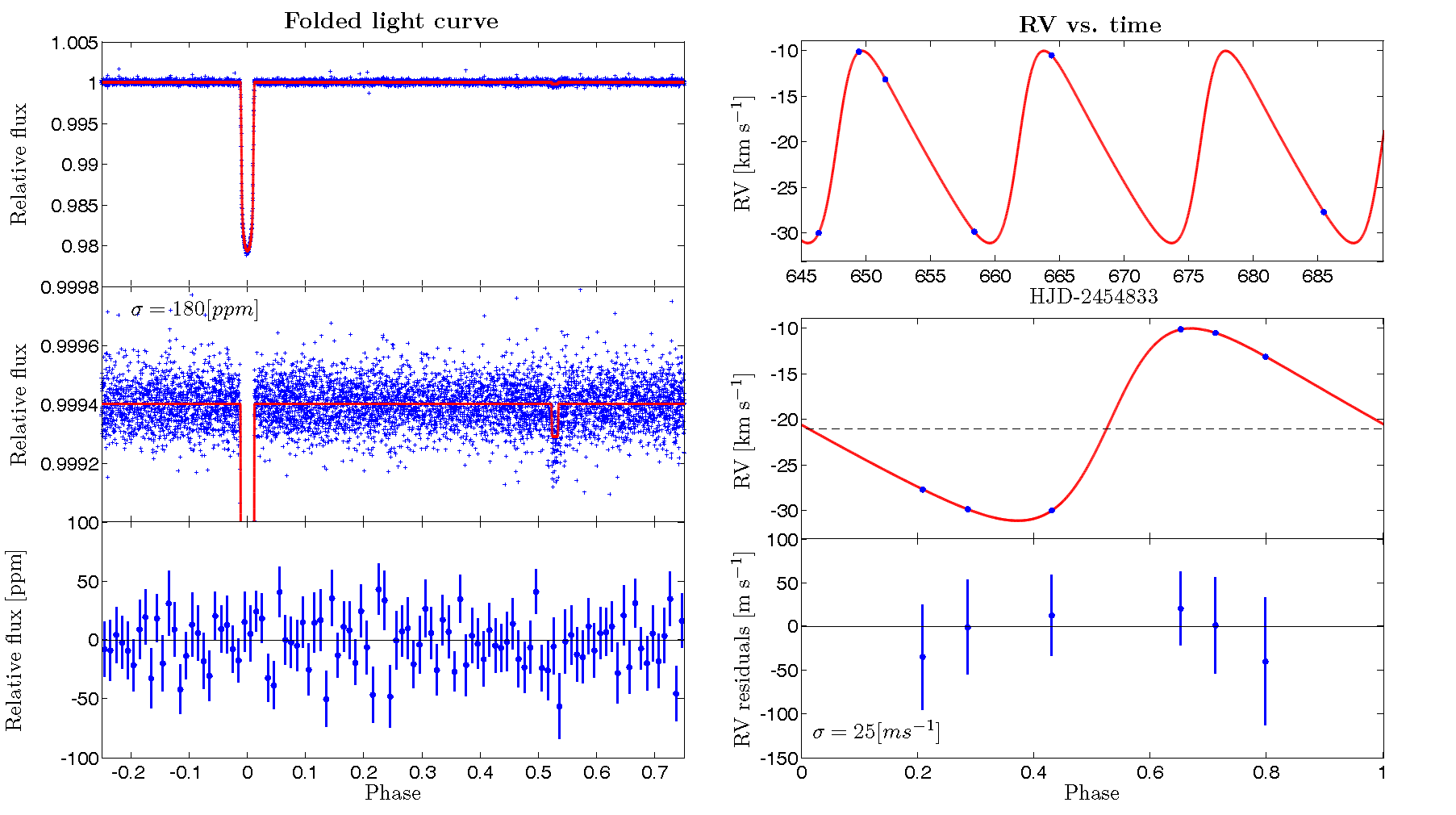}
\caption{Result of the simultaneous RV-photometry solution. {\bf Left:} the top panel shows the filtered and phased light curve (blue) and the model (red), the middle panel gives a similar but expanded view, and the bottom panel includes the model residuals vs. orbital phase binned to $1\%$ phase. {\bf Right:} the measured RVs (blue) and the model (red) are depicted vs. time (top panel), and phase (middle panel). The bottom panel gives the model residuals vs. orbital phase. Note phase zero is defined at mid transit, not at periastron passage.}
\label{ModelFigure}
\end{figure*}

\section{Physical parameters estimation}
\label{PysPar}

\begin{table}
\caption{Photospheric parameters of KIC 1571511A as derived from the spectral analysis of the co-added FIES spectrum}
\begin{tabular}{l l l l} 
\hline
\label{StellarParameters}
$T_\mathrm{eff} [^\circ$ K] 	&$\mathrm{log(g)}$ [cm s$^{-2}$] & $\mathrm{Fe/H}$ [dex] &$v \mathrm{sin} i$ [km s$^{-1}$]\\
$6195 \pm 50$		& $4.53 \pm 0.10$	&	$0.37 \pm 0.08$&	$7.9 \pm 0.5$	\\
\hline
\end{tabular}
\end{table}

Single-lined spectroscopic and eclipsing binaries, such as KIC 1571511, do not allow for the full model-free determination of their parameters. The derived quantities can only be solved up to a single line of possible mass/radius relations for each one of the components (Beatty \etal 2007, hearafter B07). One therefore needs to derive the primary's mass from some models, and systematic errors at this stage are the overwhelming source of error for this dataset (more below). We therefor took extra care and used multiple tools redundantly for the following step. We modeled the co-added FIES spectrum of 511A primarily using the new SPC fitting scheme (Buchhave \etal, in prep.) which allowed us to extract precise stellar parameters from the spectrum (see Table \ref{StellarParameters}. We double checked this new analysis with the more traditional spectral synthesis package SME (Valenti \& Piskunov 1996; Valenti \& Fischer 2005) and got a similar and  consistent results. We then used $T_\mathrm{eff}$, log(g) and Fe/H and a grid of Yonsei-Yale model isochrones (Yi \etal 2001), and performed a Monte Carlo analysis to infer the stellar mass and radius and an estimate of their uncertainties. This yielded a staller mass and radius of $M_\mathrm{511A}=1.265^{+0.036}_{-0.030} M_\mathrm{\odot}$ and $R_\mathrm{511A}=1.216^{+0.165}_{-0.043} R_\mathrm{\odot}$ respectively. Since the radius of 511A can also be inferred from the above mass and the obsrvationally-constrained relation of B07, and since the latter gives lower errors, we adopt its values: $R_\mathrm{511A}=1.343^{+0.012}_{-0.010} R_\mathrm{\odot}$. We note that at this stage we effectively have three different determinations for the primary's surface gravity: one (log(g)=4.53) comes directly from the spectral analysis, and the other two are indirect from the combination of mass and radius from the above Monte Carlo distribution (log(g)=4.37) and B07 relation (log(g)=4.28). We choose to adopt this last determination because it is least model-dependent: given the primary's mass it only assumes Kepler's laws and spherical stars.

\begin{table}
\caption{The best-fitting model of the KIC 1571511 system and derived quantities. The overwhelming error source on the physical parameters is the error on $M_1$ due to the SB1 nature of the system - so the derived errors are computed twice: with $M_1$ modeling errors included (marked with an asterisk) and without (statistical error only).}
\begin{tabular}{l l l l} 
\hline
\label{ModelResults}
Parameter 		& Value & Error & Unit\\
\hline
\hline
\multicolumn{2}{|l|}{Fitted light curve parameters} & & \\
\hline
$P$     						&   14.022480	 &  $_{-2.1}^{+2.3} \times 10^{-5}$	& d\\
$T_{mid}$     				&   4968.527088 &  $_{-9.9}^{+8.9} \times 10^{-5}$	& HJD-2450000\\
$r_1/a$						&   0.04891     &  $\pm 3.1 \times 10^{-4}$	& \\
$r_2/r_1$     				&   0.13277     &  $_{-4.6}^{+3.8} \times 10^{-4}$	& \\
$i$      					&  89.480       &  $_{-0.056}^{+0.069}$		& degrees\\
$e cos(\omega)$			&  -0.04057     &  $\pm 4.0 \times 10^{-4}$	& \\
$e sin(\omega)$			&   0.3244      &  $_{-2.6}^{+2.8} \times 10^{-3}$	& \\
$L_2/L_{tot}$  			&  $2.75 \times 10^{-4}$ & $\pm 0.19 \times 10^{-4}$	& \\
$u_{1,1}$     				&   0.373       &  $\pm 0.019$               & \\
$u_{2,1}$     				&   0.205       &  $_{-0.045}^{+0.047}$ 		& \\
$u_{1,2}$     				&   0.43        &  $_{-0.30}^{+0.38}$			& \\
$q=\frac{m_2}{m_1+m_2}$	&	 0.10052     &  $\pm 2.3 \times 10^{-4}$	& \\
$L_3/L_{tot}$				&	 0				 &  (fixed)							& \\
\hline
\multicolumn{2}{|l|}{Fitted RV parameters} & & \\
\hline
$\gamma$     				&  -21030.2     &  $\pm 3.7$						& m s$^{-1}$ \\
Scale     					&   1.265	 	 &  (fixed) 						& $M_\odot$ ($M_1$ model)\\
\hline
\multicolumn{2}{|l|}{Derived parameters} & & \\
\hline
K     					   &  10521        &  $\pm 24$						& m s$^{-1}$ \\
e     					   &  0.3269       &  $\pm 0.0027$ 					& \\
b								&  0.383			 &	 $_{-0.049}^{+0.040}$		& \\
$\omega_1$					&  82.872       &  $\pm 0.099$ 					& degrees \\
$\rho_1$ 					&  740 			 &  $\pm 14$ 						&$\mathrm{kg \ m^{-3}}$\\
					 			&  0.5242		 &  $\pm 9.9 \times 10^{-3}$	& $\rho_\odot$\\
$\mathrm{log \ g_2}$ 	&  5.0875 		 &  $_{-7.6}^{+8.0} \times 10^{-3}$	& cm {sec}$^{-2}$ \\
Mass function				&  $142.8 \times 10^{-5}$	&  $\pm 1.02 \times 10^{-5}$ 	& $M_\odot$ \\
\hline
\multicolumn{2}{|l|}{Physical parameters} & & \\
\hline
$M_1$ 						&  1.265			 &  $_{-0.030}^{+0.036}$ 			& $M_\odot$ (from model)\\
$M_2$ 						&  0.14136		 &  $\pm 3.6 \times 10^{-4}$	& $M_\odot$\\
								&  		 		 &  $_{-42}^{+51} \times 10^{-4}$& $M_\odot$ * \\
$R_1$ 						&  1.343        &  $_{0.010}^{+0.012}$				& $R_\odot$\\
$R_2$ 						&  0.17831		 &  $_{-6.2}^{+5.1}\times 10^{-4}$& $R_\odot$\\
								& 					 &  $_{-16}^{+13} 10^{-4}$ 		& $R_\odot$ *\\
$T_\mathrm{eff,511B}$	& 4030-4150		 &  (see text)							& $^\circ K$\\
\hline
\end{tabular}
\end{table}

511B is thus determined to have a mass of $0.14136 \pm 3.6 \times 10^{-4} ; \: _{-42}^{+51} \times 10^{-4} M_\odot$ and a radius of $0.17831 _{-6.2}^{+5.1} \times 10^{-4} ; \: _{-16}^{+13} \times 10^{-4} R_\odot$ (statistical and modeling errors, respectively). Not surprisingly, both the LC and the RV data are far more precise than the model-dependent derivation of the mass of 511A, and the latter is the overwhelming source of error in the derived physical parameters of 511B. We therefore quote two error estimations of the mass and radius of 511B in Table \ref{ModelResults} -- the larger is derived only from the modeling error of the mass of 511A, and the smaller is derived only from statistical errors of the LC-RV fit.

Since we did not model the reflection effect the $L_2/L_{tot}=2.75 \times 10^{-4}$ depth of the secondary eclipse is a combination of the primary's flux reflected off the secondary, and of the intrinsic luminosity of 511B itself. The former can be calculated as $F_{reflected} = A_g(\frac{r_2}{a})^2 sin(i)$ where $A_g$ is the geometric albedo and $a$ is the primary-secondary distance during secondary eclipse, which in this case is: $F_{reflected} = A_g \times 4.2 \times 10^{-5}$. We find that reflected light cannot contribute more than about 1/6 of the light lost during the secondary eclipse. Thus at least $2.33 \times 10^{-4}$ (and probably much closer to the entire $2.75 \times 10^{-4}$) of the total flux in the {\it Kepler} passband can be attributed to 511B's intrinsic luminosity. A toy model\footnote{see Kepler Instrument Handbook, document KSCI-19033, for full description.} (using a uniform {\it Kepler} passband between 420 and 900 nm and blackbody spectral densities) gives a temperature range for 511B of $T_{eff,511B}=4030^\circ K$ to $4150^\circ K$ for the above contrast range.

\section{Discussion}
\label{Discussion}

We present the initial characterization of {\it Kepler} EB KIC 1571511. We show that the secondary of this EB is a very low mass star with a mass of $0.14136 M_\odot$ and a radius of $0.17831 R_\odot$ - so its diameter is smaller that some planet (e.g. Hartman \etal 2011, Anderson \etal 2011b). For a low-mass object to be considered for a "benchmark status" of its class one would want that the object would be physically associated with a more Sun-like star, since such stars are currently better understood. Better still are such binaries that are eclipsing, and the best constraints could come from fully-eclipsing and double-lined such binaries, where masses and radii are arrived at model-free. KIC 1571511 is almost such an EB -- currently lacking only the secondary's RV signal. Indeed, if one plots the uncertainty in mass and radius of all well-characterized (both errors under 5\%) low-mass objects and of 511B (Figure \ref{RandMprecision}), it is easy to see that the latter occupies a unique spot of the lowest-mass well characterized star just above the brown-dwarfs to stars transition (0.075 $M_{Sun}$). Importantly, the overwhelming source of error on both the mass and radius of 511B is the error on the mass of the primary, 511A. This is due, on the one hand, to the single-lined nature of the system in the visible band, and, on the other hand, to the exquisite quality of the {\it Kepler} LC which allows for very precise determination of all LC-derived quantities. This, in turn, means that the continued {\it Kepler} observations on the target guarantee marked improvements in all LC-derived quantities in the future, while any observation of the system as an SB2 (perhaps in the infrared) will dramatically reduce the overall error on both the mass and radius of 511B. This is visualized with the empty symbol of 511B on Figure \ref{RandMprecision} which show that discounting the (modeling) error on the mass of 511A drastically reduces the errors on the parameters of 511B to potentially the best characterized and lowest-mass object -- so a potential benchmark object indeed.

\begin{figure}
\includegraphics[width=84mm]{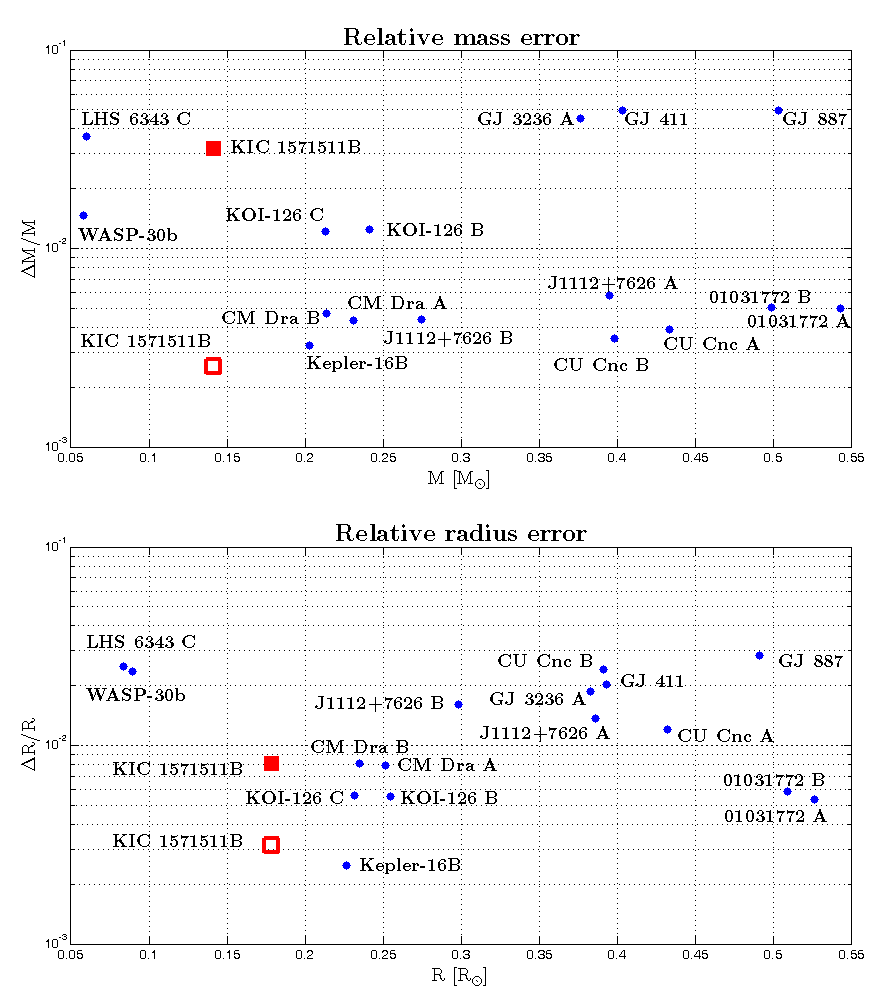}
\caption{Relative precision of the mass (upper panel) and radius (lower panel) of all well-characterized low-mass stars ($\mathrm{M_\star < 0.55 M_\odot}$) with both determined to better than $5\%$ (blue circles). Also noted (red square) is the location of KIC 1571511B - once derived only from the error on the mass of 511A (filled symbol) and once derived only from statistical errors (empty symbol). References: LHS 6343 C (Johnson \etal 2011), WASP-30b (Anderson \etal 2011a), KOI-126 B,C (Carter \etal 2011), CM Dra A \& B, CU Cnc A \& B, GJ 411, GJ 887 (L{\'o}pez-Morales 2007 and references therein), GJ 3236 A (Irwin \etal 2009), LSPM J1112+7626 A and B (Irwin \etal 2011), Kepler-16B (Doyle \etal 2011)}
\label{RandMprecision}
\end{figure}

When compared with other known low-mass stars all the other systems we encountered are less favorable to serving as benchmarks of this kind: Most known late M dwarfs and BDs are not in eclipsing systems at all. Nearly all low-mass EBs, and especially those observed from the ground, have LCs that cannot compete with {\it Kepler}'s exquisite quality. However, there are a number other interesting objects already in that dataset: KOI-126 B,C (Carter \etal 2011) and Kepler-16B (Doyle \etal 2011) are indeed very low mass stars -- but all are part of compact hierarchical triple systems, making follow up and analysis more difficult. We conclude that KIC 1571511B is indeed uniquely situated to become a benchmark object for the study of low-mass stars.


\label{lastpage}


\begin{thebibliography}{99}

\bibitem[Anderson \etal (2011)]{2011ApJ...726L..19A} Anderson, D.~R., \etal \ 2011a, \apjl, 726, L19 
\bibitem[Anderson \etal (2011)]{2011MNRAS.416.2108A} Anderson, D.~R., Smith, A.~M.~S., Lanotte, A.~A., et al.\ 2011b, \mnras, 416, 2108 

\bibitem[Beatty \etal (2007)]{2007ApJ...663..573B} Beatty, T.~G., \etal \ 2007, \apj, 663, 573 

\bibitem[Borucki \& for the Kepler Team(2010)]{2010arXiv1006.2799B} Borucki, W.~J., \& for the Kepler Team 2010, arXiv:1006.2799 

\bibitem[Buchhave \etal (2010)]{2010ApJ...720.1118B} Buchhave, L.~A., \etal \ 2010, \apj, 720, 1118 

\bibitem[Carter \etal (2011)]{2011Sci...331..562C} Carter, J.~A., \etal \ 2011, Science, 331, 562 

\bibitem[Coughlin \etal (2011)]{2011AJ....141...78C} Coughlin, J.~L., L{\'o}pez-Morales, M., Harrison, T.~E., Ule, N., \& Hoffman, D.~I.\ 2011, \aj, 141, 78 

\bibitem[Doyle \etal (2011)]{2011Sci...333.1602D} Doyle, L.~R., Carter, 
J.~A., Fabrycky, D.~C., \etal \ 2011, Science, 333, 1602 

\bibitem[Enoch \etal (2010)]{2010A&A...516A..33E} Enoch, B., Collier Cameron, A., Parley, N.~R., \& Hebb, L.\ 2010, \aap, 516, A33 

\bibitem[Hartman \etal (2011)]{2011arXiv1106.1212H} Hartman, J.~D., Bakos, G.~{\'A}., Torres, G., et al.\ 2011, arXiv:1106.1212 

\bibitem[Irwin \etal (2009)]{2009ApJ...701.1436I} Irwin, J., \etal \ 2009, 

\bibitem[Irwin \etal (2011)]{2011arXiv1109.2055I} Irwin, J.~M., Quinn, S.~N., Berta, Z.~K., \etal \ 2011, arXiv:1109.2055  

\bibitem[Johnson \etal (2011)]{2011ApJ...730...79J} Johnson, J.~A., \etal \ 2011, \apj, 730, 79 

\bibitem[Kipping(2010)]{2010MNRAS.408.1758K} Kipping, D.~M.\ 2010, \mnras, 408, 1758 

\bibitem[Lacy(1977)]{1977ApJS...34..479L} Lacy, C.~H.\ 1977, \apjs, 34, 479 

\bibitem[L{\'o}pez-Morales(2007)]{2007ApJ...660..732L} L{\'o}pez-Morales, M.\ 2007, \apj, 660, 732 

\bibitem[Mandel \& Agol(2002)]{2002ApJ...580L.171M} Mandel, K., \& Agol, E.\ 2002, \apjl, 580, L171 

\bibitem[Ribas(2006)]{2006Ap&SS.304...89R} Ribas, I.\ 2006, \apss, 304, 89

\bibitem[Southworth \etal (2004)]{2004MNRAS.351.1277S} Southworth, J., Maxted, P.~F.~L., \& Smalley, B.\ 2004a, \mnras, 351, 1277 

\bibitem[Southworth \etal (2004)]{2004MNRAS.355..986S} Southworth, J., Zucker, S., Maxted, P.~F.~L., \& Smalley, B.\ 2004b, \mnras, 355, 986 

\bibitem[Torres \& Ribas(2002)]{2002ApJ...567.1140T} Torres, G., \& Ribas, I.\ 2002, \apj, 567, 1140 

\bibitem[Tegmark \etal (2004)]{2004PhRvD..69j3501T} Tegmark, M., \etal \ 2004, \prd, 69, 103501 

\bibitem[Udry \etal (1999)]{1999ASPC..185..367U} Udry, S., Mayor, M., \& Queloz, D.\ 1999, IAU Colloq.~170: Precise Stellar Radial Velocities, 185, 367 

\bibitem[Valenti \& Piskunov(1996)]{1996A&AS..118..595V} Valenti, J.~A., \& Piskunov, N.\ 1996, \aaps, 118, 595 

\bibitem[Zucker \etal (2007)]{2007ApJ...670.1326Z} Zucker, S., Mazeh, T., 
\& Alexander, T.\ 2007, \apj, 670, 1326 

\end{thebibliography}
\end{document}